\documentclass[10pt]{article}

\usepackage{amsmath}
\usepackage{amssymb}

\usepackage{graphicx}

\usepackage{cite}

\usepackage{color} 

\usepackage[labelfont=bf,labelsep=period,justification=raggedright]{caption}

\makeatletter
\renewcommand{\@biblabel}[1]{\quad#1.}
\makeatother

\date{}

\begin{document}
\vspace*{0.35in}

% Title must be 250 characters or less.
% Please capitalize all terms in the title except conjunctions, prepositions, and articles.
\begin{flushleft}
{\Large
\textbf\newline{Benchmarking inverse statistical approaches for protein structure and design with exactly solvable models}
}
\newline
% Insert author names, affiliations and corresponding author email (do not include titles, positions, or degrees).
\\
Hugo Jacquin  \textsuperscript{1}, 
Amy Gilson  \textsuperscript{2}, 
Eugene Shakhnovich  \textsuperscript{2}, 
Simona Cocco \textsuperscript{1,$\dagger$},
R\'emi Monasson \textsuperscript{3,$\dagger$}
\\
\bigskip
\bf{1} Laboratory of Statistical  Physics, Ecole Normale Sup\'erieure, CNRS, PSL Research University,  Sorbonne Universit\'es UPMC, 24 rue Lhomond, 75005 Paris, France\\
\bf{2} Department of Chemistry and Chemical Biology, Harvard University, 12 Oxford Street, Cambridge, MA 02138, USA\\
\bf{3}  Laboratory of Theoretical Physics, Ecole Normale Sup\'erieure, CNRS, PSL Research University,  Sorbonne Universit\'es UPMC, 24 rue Lhomond, 75005 Paris, France
\\
\bigskip

% Insert additional author notes using the symbols described below. Insert symbol callouts after author names as necessary.
% 
% Remove or comment out the author notes below if they aren't used.
%
% Primary Equal Contribution Note
%\Yinyang These authors contributed equally to this work.

% Additional Equal Contribution Note
% Also use this double-dagger symbol for special authorship notes, such as senior authorship.
%\ddag These authors also contributed equally to this work.

% Current address notes
%\textcurrency a Insert current address of first author with an address update
% \textcurrency b Insert current address of second author with an address update
% \textcurrency c Insert current address of third author with an address update

% Use the asterisk to denote corresponding authorship and provide email address in note below.
*Corresponding author: monasson@lpt.ens.fr; $\dagger$ contributed equally as last authors

\end{flushleft}
% Please keep the abstract below 300 words
\section*{Abstract}
Inverse statistical approaches to determine protein structure and function from Multiple Sequence Alignments (MSA) are emerging as powerful tools in computational biology. However the underlying assumptions of the relationship between the inferred effective Potts Hamiltonian and real protein structure and energetics remain untested so far. Here we use lattice protein model (LP) to benchmark those inverse statistical approaches. We build MSA of highly stable sequences in target LP structures, and infer the effective pairwise Potts Hamiltonians from those MSA. We find that inferred Potts Hamiltonians reproduce many important aspects of `true' LP structures and energetics. Careful analysis reveals
that effective pairwise couplings in inferred Potts Hamiltonians depend not only on the energetics of the native structure but also on competing folds; in particular, the coupling values reflect both positive design (stabilization of native conformation) and negative design (destabilization of competing folds). In addition to providing detailed structural information, the inferred Potts models used as protein Hamiltonian for design of new sequences are able to generate with high probability completely new sequences with the desired folds, which is not possible using independent-site models. Those are remarkable results as the effective LP Hamiltonians used to generate MSA are not simple pairwise models due to the competition between the folds. Our findings elucidate the reasons for the success of inverse approaches to the modelling of proteins from sequence data, and their limitations.

\section*{Author Summary}
Inverse statistical approaches, modeling pairwise correlations between amino acids in the sequences of homologous proteins across many different organisms, can successfully extract protein structure (contact) information. Here, we benchmark those statistical approaches on exactly solvable models of proteins, folding on a 3D lattice, to assess the reasons underlying their success and their limitations. We show that the inferred parameters (effective pairwise interactions) of the statistical models have clear and quantitative interpretations in terms of positive (favoring the native fold) and negative (disfavoring competing folds) protein sequence design. New sequences randomly drawn from the statistical models are likely to fold into the native structures when effective pairwise interactions are accurately inferred, a performance which cannot be achieved with independent-site models.

 %\linenumbers

\section*{Introduction}

\vskip .3cm \noindent
Prediction of protein structure from sequence remains a major goal of computational structural biology with significant practical implications. While great progress was achieved in de novo prediction of structures and even direct folding of smaller globular water soluble proteins \cite{baker01,Shaw2011}, structure prediction remains  challenging for larger proteins, and membrane and other non-globular proteins. For these cases indirect methods such as homology modeling are most promising.  An approach to predict structure from the statistics of multiple sequence alignment (MSA)  was proposed twenty years ago \cite{gobel94}. The underlying assumption for these statistical approaches is that residues that covary in a MSA are likely to be in close proximity in protein structure \cite{neher94}. 

Recently, this approach has been significantly improved and has become a practical tool to extract structural information from sequence data \cite{Weigt:2009ba,Morcos:2011c}, and to help in folding proteins \cite{Hopf:2012kv}.  Progress was made possible by the application of the maximum entropy principle of statistical mechanics \cite{jaynes57a,Bialek2012_book} to derive the distribution of sequences in a protein family, {\em i.e.} their probability of appearing in an MSA. This approach is similar in spirit to the derivation of Gibbs distribution in Statistical Mechanics, with an effective Hamiltonian constructed to reproduce single-site amino-acid frequencies and pairwise amino-acid correlations. As such, the effective Hamiltonian is a Potts model \cite{wu82} with parameters (the site-dependent fields and pairwise couplings acting on amino acids) fitted to reproduce the MSA statistics. This approach allows one to disentangle direct  (corresponding to couplings) from indirect (mediated by other sites) correlations \cite{Morcos:2011c,burger10}; large couplings are much better predictors of contacts than large correlations.

The exploitation of covariation information in proteins extends beyond structural prediction, and is potentially useful for homology detection, for characterizing the fitness changes resulting from mutations, or for designing artificial proteins with `natural' properties \cite{valentia2013}. Experiments show in particular that a sizeable fraction of artificial sequences generated to respect the 1- and 2-point amino-acid frequencies calculated from the natural MSA of the WW domain (a short protein domain with $\simeq $30 amino acids) acquire the native fold \cite{Socolich:2005js,russ05}. This fraction vanishes when artificial sequences are generated which respect the pattern of single-site frequencies only. Recent studies have shown that inferred maximum-entropy models are helpful to predict the effects of mutations in various protein sequences \cite{Lui13,Morcos14rt,Figliuzzi15}, including HIV virus sequences \cite{Mann:2014ji}. 

Despite those successes fundamental issues remain poorly understood.  To what extent do the inferred couplings reflect the physical  (energetics)  and structural (contact) properties of proteins? How much are the covariation properties of one family influenced by the need to prevent sequences from folding into competing structures? Are pairwise correlations, and hence, couplings generally sufficient to capture the statistical features of protein families and define generative models? 

In this work we answer those questions using lattice-based protein (LP) models \cite{gutin90,shakh93,li1996}.  Using LP allows us to evaluate the strengths  and limitation of the Potts inference methods by comparing recovered solutions with known exact ones. Protein-like sequences are generated using a well-defined Hamiltonian that employs 20 amino-acid types and the Miyazawa-Jernigan energy function to approximate contact energies \cite{jernigan85}. LP share many common properties with real proteins, for instance nontrivial statistical features of the sequences associated with a given fold, with variable conservation along the protein chain \cite{shakh93,shakhnovich96}.  Surprisingly, the covariation properties of LP had not been much studied so far \cite{noivirt09}. Here, we do so by generating MSA data for various LP folds, and apply to those sequence data the same inverse approaches used for real proteins  (Fig.~1).  In particular, we show that the inferred couplings are excellent predictors of contacts in the protein structure. In addition, while the inferred Potts pairwise couplings mostly reflect the energetics and the contacts in the fold corresponding to the MSA, they also strongly depend on the nature of the other folds competing with the native structure, and have transparent interpretations in terms of positive and negative designs. Furthermore, we show that the pairwise Potts model is generative: it produces with high probability sequences with the right fold, a performance which cannot be achieved by models reproducing single-site frequencies alone. This is a non-trivial result as the log probability of LP sequences constrained to fold in a given structure is not a sum of pairwise contributions only, but includes multi-body interactions to all orders due to contributions from competing folds.

 % Results and Discussion can be combined.
\section*{Results}

\vskip .3cm\noindent We have generated MSA for four representative LP structures shown in Fig.~2A, referred to as $S_A$ to $S_D$; the energetics of the LP model, the Boltzmann probability $P_{nat}(S|{\bf A})$ of a sequence $\bf A$ to achieve a fold $S$, and the procedure for sampling the sequence space and generate MSA are described in the Methods section.  Any structure $S$ is characterized by the volume it occupies in sequence space (Fig.~1), that is, by the number of sequences $\bf A$ it is associated with, {\em e.g.} with values of $P_{nat}(S|{\bf A})$ larger than, say, 0.995. For the four structures considered here, the numbers of associated sequences are extremely large compared to the MSA we have generated ($5\, 10^4$ sequences). The higher this number, the more designable the corresponding structure is expected to be \cite{li1996}.  The designability of each structure can be empirically  estimated  from the diversity of the sequences in the corresponding MSA, from the spectral properties of the contact matrix \cite{england2003}, and, as shown below in this paper, from the entropy of the Potts model inferred from the MSA (Methods and Table~1). More information about how to define and estimate designability can be found in S1 Text, Section I.B.

\vskip .3cm
\noindent{\bf Potts pairwise couplings give accurate contact prediction}  

\vskip .3cm \noindent
We show in Fig.~2A the positive predictive value (PPV) for contact prediction (Methods) for the four structures $S_A,S_B,S_C,S_D$, based on the ranking of the mutual information (MI) scores \cite{chiu91} and of the inferred Potts couplings, with the mean field (DCA) \cite{Morcos:2011c}, the pseudo likelihood (PLM) \cite{ekeberg12}, and the adaptive cluster expansion (ACE) \cite{Cocco:2011fo,Cocco:2012id,Bar:2015} procedures. A fifth method, called Projection, shown with magenta lines in Fig.~2A will be introduced later on. Mean-field DCA is a very fast, approximate method to infer the couplings. PLM is another approximation method, known to be remarkably accurate for contact prediction on real protein data. ACE is slower, but provides a more precise estimate of the coupling parameters (Methods). As in the case of real protein data, Potts-based contact predictions, either with DCA, PLM or ACE, generally outperform MI-based predictions. MI, indeed, does not disentangle large indirect correlations mediated by one or more sites from direct correlations due to contacts. PLM and ACE  couplings are more precise than their DCA counterparts and accordingly, give better contact predictions for all four structures (Table~2).  As shown later, most of the missed contacts with ACE (3 or 4, depending on the structure in Fig.~2A) are carried by the site at the center of the structure. Why this is so and how to improve contact prediction (magenta lines in Fig.~2A) will become clear from the detailed interpretation of the Potts couplings below. 

While the results above were obtained for uniformly sampled MSAs, real protein data suffer from imperfect sampling, {\em e.g.} resulting from phylogenetic correlations. To analyze the effects of biases in the sampling on structural predictions in a simple and controlled way we have introduced a sampling procedure that favors sequences with small Hamming distances to a reference, wild-type sequence (Methods). As shown in Fig.~2B, modifying the strength of the bias, $b$, allows us to interpolate smoothly between the uniform sampling above ($b=0$) and strongly biased MSAs with reduced sequence diversities (large $b$ values). Results for contact prediction obtained with the PLM procedure are reported in Fig.~2B. We observe that performances worsen as the bias in the MSAs increases. Empirical reweighting of the sequence statistics according to their similarity with the other sequences in the MSA (Methods), as usually done for real protein data, efficiently improves the quality of the structural prediction for intermediate bias values. For very strong biases ($b=0.1$ in Fig.~2B), the effective number of sequences resulting from reweighting is so low that results become very poor.

\vskip 1cm
\noindent{\bf Potts model generates good folding sequences with high probability, in contrast to Independent-site Model}  

\vskip .3cm
\noindent We now test the ability of the inferred Potts model to be generative, that is, to produce sequences having high probability of folding into the native structure. To do so, we infer the pairwise Potts model, Eq.~[\ref{Potts}], from the MSA, hereafter referred to as `natural',  of structure $S_B$ with the ACE procedure. We sample the Potts distribution using Monte Carlo simulations, thereby generating a new MSA, hereafter referred  to as `Potts-ACE'.  We then compute the probabilities of folding (into $S_B$) of all the sequences in the Potts-ACE MSA,  see Eq.~[\ref{pnat}]. Results are shown in Fig.~3A. Strikingly, the `majority of the sequences have high folding probabilities. Conversely, sequences generated with an Independent-site Model (IM, see Methods) following the same procedure are unlikely to have the right fold, as shown in Fig.~3A. Pairwise interactions are therefore crucial to design new sequences with the right fold. Because the Potts-ACE model is generative, its entropy can be used as a quantitative estimate of the designability of the structure, with results reported in Table~1 and S1 Text, Section I.B.

To better characterize the generative properties of Potts-ACE, we show in Fig.~3B the scatter plot of the `energies'  ${\cal H}^{Potts-ACE}[{\bf A}; {\bf h},{\bf J}]$, Eq.~[\ref{Potts}], with the inferred Potts-ACE model vs. the effective Hamiltonian $-\beta \log P_{nat}(S_B|{\bf A})$ for the best folding sequences $\bf A$ in the MSA generated with the Potts-ACE model for structure $S_B$. We observe a roughly monotonic relationship between the two quantities: the lower the energy with the Potts-ACE model, the more likely is the sequence to be a good folder (Fig.~3B). However, while the Potts-ACE model rightly predicts that sequences with increasing energies are less likely to be good folders, the value of ${\cal H}^{Potts-ACE}$ is less and less predictive about the precise value of $P_{nat}$, as shown by the increasing dispersion in Fig.~3B. Note that the low-energy part of this scatter plot, corresponding to the best sequences generated with the Potts-ACE model, is quantitatively similar to its counterpart for the `natural' MSA, see Fig.~F in S1 Text.

We stress that Potts couplings have to be inferred with high precision to generate high-quality sequences. We show in Fig.~3A that sequences generated by the Potts-Gaussian model, which makes use of the approximate DCA couplings (details in S1 Text, Section II.B), have very high folding probabilities, but are extremely concentrated around the consensus sequence. The Potts-Gaussian model, contrary to the IM and the Potts-ACE model, fails to reproduce the diversity of sequences observed in the natural MSA. This failure is a direct consequence of the Gaussian approximation (S1 Text, Section~IV). The Potts-PLM model reproduces the diversity of natural MSA, but generates sequences, whose folding properties largely vary from very good to very poor (Fig.~3A). The poor folding discrimination is mostly due to the large regularization parameter used in the inference (S1 Text, Section II.C and Fig.~B). In summary,  the Potts-ACE model, in contradistinction with IM, DCA and PLM, is capable to generate a large set of diverse sequences that fold with high probability in the target native structure.

\vskip .3cm
\noindent{\bf Potts couplings reflect both energetics in the native fold and competition with other folds} 

\vskip .3cm \noindent
We now study in more detail the properties of the Potts couplings.  We show in Fig.~4A the scatter plot of the Potts couplings  $J_{ij}(a,b)$ inferred with the ACE method for structure $S_B$ vs. the Miyazawa-Jernigan energetics parameters $-E(a,b)$ used to compute the LP energies (Methods); scatter plots for the other structures are shown in Text S1, Figs.~H, I \& J. For each pair of sites $i,j$, we observe a linear  dependency,
\begin{equation}
J_{ij}(a,b) \approx \lambda_{ij}\;\times\; \big( - E(a,b)\big)  \ ,
\label{slopedef}
\end{equation}
where the slope $\lambda_{ij}$  is positive  for the pairs in contact in the native structure (red symbols in Fig.~4A), and negative, or zero  for the pairs not in contact (blue symbols). In the following, we characterize each pair $i,j$ with its slope, computed through
\begin{equation}
\lambda_{ij} = \frac{- \sum_{ab} J_{ij}(a,b) E(a,b)}{ \sum_{ab} E(a,b)^2} \  .
\label{scoring_projMJ}
\end{equation}

We interpret $\lambda_{ij}$ as a measure of the coevolutionary pressure on the sites $i,j$, due to the design of the native structure. This interpretation is supported by the following theoretical and approximate expression for the pressure (see Methods and derivation in S1 Text, Section~III):
\begin{equation}
\lambda^{TH}_{ij}\approx \frac{c_{ij} -\bar c_{ij}} {\log\left[1+\frac{1}{\beta\,N_S \,e^{-\Delta}} \right]} \ .
\label{slope}
\end{equation}
The numerator, $ c_{ij}-\bar c_{ij}$, measures the difference between the native contact map, $c_{ij}$ ($=1$ if  $i,j$ are in contact and 0 otherwise), and the average contact map,  $\bar c_{ij}$, of the structures in competition with the native fold, each weighted by $e^{-\Delta}$, where $\Delta$ is the typical gap between the energy of sequences folded into the native structure and their energies in the competing structures (Methods) \cite{sali94}. The native and average contact maps for structure $S_B$ are shown, respectively, in the lower and upper triangles of Fig.~4B. The denominator in Eq.~[\ref{slope}] depends on the inverse sampling temperature, $\beta$, in the Monte Carlo procedure used to generate the MSA, the effective number, $N_S$, of competing structures, and the typical energy gap, $\Delta$ (Methods and S1 Text, Section~III.D). 
%The pressure $\lambda$ is an increasing function of $\beta$ and of the amount of competition from the other structures, measured by $N_S\, e^{-\Delta}$, in agreement with  Eq.~[\ref{slope}].

Figure~4C shows the pressures, $\lambda_{ij}$, computed from the inferred couplings, Eq.~[\ref{scoring_projMJ}], vs. the contact-map difference, $\delta c_{ij} \equiv c_{ij}-\bar c_{ij}$, across all pairs of sites $i,j$ for the native fold $S_B$. We observe a monotonic dependence of the pressure $\lambda_{ij}$ with $\delta c_{ij}$; in particular, $\lambda_{ij}$ has the same sign as $\delta c_{ij}$, expect for a few, small (in absolute value) $\lambda_{ij}$ and $\delta c_{ij}$. Note that pairs $i,j$ such that $j-i$ is even valued can never be in contact in any structure on a cubic lattice: $c_{ij}=\bar c_{ij}=0$, as can be seen in the checkerboard pattern of $\bar c_{ij}$ in Fig.~4B. The associated pressures,  $\lambda_{ij}$, are weak and indicate the level of noise in the inference (magenta symbols in Fig.~4C). 

A detailed classification of the pairs of sites is shown in Fig.~4C, based on the identification of the closest competitor to $S_B$, structure $S_F$, represented in Fig.~4D. The closest competitor is defined as the structure that is closest in energy (smallest gap) to the native structure, see Methods, Eq.~[9]. Informally speaking, $S_F$ is the `most' dangerous structure as sequences folding in $S_B$ could fold in $S_F$ upon a few undesirable mutations. As the native fold is not in competition with a single structure, the notion of closest competitor is somewhat reductive, but is helpful to understand the mechanisms controlling covariation and the meaning of the inferred couplings. In particular, comparison of the contact maps of $S_B$ and $S_F$ allows us to define four classes of contacts, and to illustrate concretely the mechanisms for positive and negative designs \cite{bere07,noivirt09}. 

Pairs with large and positive $\delta c_{ij}$ are in contact in the native fold, but not in the competitor structures. They are under strong covariation pressure to stabilize the native fold and not the competitors (positive design).  We see in Fig.~4C that the pairs of sites in contact in $S_B$ and not in $S_F$ (class Unique Native, UN) are associated with large couplings. On the contrary, pairs with small and positive $\delta c_{ij}$ are in contact both in the native and competitor folds, and are subject to weak pressures. As a result, the corresponding couplings are weak, and contact prediction is difficult. Pairs of sites in contact in both $S_B$ and its closest competitor, $S_F$, form the Shared Native (SN) class in Fig.~4C. For the structures we have studied, those contacts are generally established between the site at the center of the cube and its four neighbors  (see fold $S_B$ in Fig.~2A and its contact map in Fig.~4B) as these central contacts are typically shared with closest competitor folds (Fig.~4D and Fig.~K in S1 Text). 

Sites $i,j$ not in contact in the native fold correspond to negative $\delta c_{ij}$. Most of those pairs are not in contact in the competitor structures either and correspond to weak pressures and weak inferred couplings. This phenomenon is illustrated by the Non-Native (NN) class, which gathers the pairs of sites that are neither in contact in $S_B$ more in $S_F$, see Fig.~4C.  However, some pairs do correspond to contacts in the competitor folds (large and negative $\delta c_{ij}$), and are subject to strong and negative pressures.  The anti-correlation between the couplings  $J_{ij}(a,b)$ and the Miyazawa-Jernigan parameters $E(a,b)$ is the result of negative design: the native structure is favored by rendering contacts that appear only in competitor folds unfavorable \cite{bere07}. Negative design is clearly visible for the pairs of sites in the Close Competitor (CC) class in Fig.~4C, which are in contact in $S_F$ but not in the native fold, $S_B$. 

As the scores used for contact predictions are based on the squared couplings (Methods), pairs of sites in the CC class, associated with large couplings (anti-correlated with the Miyazawa-Jernigan energy parameters), may have larger scores than true contacts in the SN class, associated with weaker couplings (positively correlated with MJ). To avoid those false positives, we rank pairs of sites according to their projection on the MJ matrix, defined in Eq.~[\ref{scoring_projMJ}] (Methods). This ranking method, referred to as  'projection',  improves contact prediction, see PPV in magenta in Fig.~2A. It allows us to detect, in particular, the contacts of the central site, which are often shared among the native and competing structures (see contact maps in Fig.~4B for structure $S_B$ and its competitors), and which correspond to weak, positive pressures. We find, for the four structures studied, that 27 out of the 28 largest projections do correspond to real contacts (Table~2). 

The evolutionary pressures, averaged over the pairs of sites in the UN, SN and CC classes, give $\lambda_{UN}\approx 2.8$,   $\lambda_{SN}\approx 1.4$,   $\lambda_{CC}\approx -0.9$. Those results are comparable to, but smaller than, the average theoretical predictions from Eq.~[\ref{slope}]: $\lambda^{TH}_{UN}=4.5$,  $\lambda^{TH}_{SN}=2.7$ and  $ \lambda^{TH}_{CC}=-1.8$. Larger values, closer to the theoretical estimates are obtained when the regularization parameter, penalizing large couplings, is made smaller (Methods). To further test the theoretical prediction of the pressure, Eq.~[\ref{slope}], we have varied two features of the Monte Carlo procedure used to generate the MSA: (1) the pool of possible competing structures appearing in the denominator of the folding probability, $P_{nat}$, in Eq.~[\ref{pnat}], and (2) the inverse sampling temperature $\beta$ (Methods). We observe that the pressures $\lambda_{ij}$ increase when the pool of competing structures is restricted to structures similar to the native one, or when $\beta$ is increased, as sequences are constrained to have higher values for $P_{nat}$  (Fig.~L and Table~A in S1 Text).

\section*{Discussion}

\vskip .5cm \noindent
Lattice proteins offer a fully controlled, nontrivial benchmark to characterize the factors that affect the success of inverse statistical approaches in predicting structure from multi sequence alignments (MSA). The maximum entropy Potts model, inferred to reproduce the low-order statistics of an MSA, can be successfully used for structure prediction. Potts-based predictions, in particular with accurate sophisticated inference methods (PLM, ACE), outperform Mutual Information approaches on lattice protein data (Fig.~2A), as for real protein data. An important finding in the present work is the approximate linear relationship between the inferred Potts couplings, $J_{ij}(a,b)$, and the Miyazawa-Jernigan energetic parameters, $E(a,b)$, across the pairs of amino acids $a,b$ for a given pair of sites $i,j$, Eq.~[\ref{slopedef}] and Fig.~4A. The slope of this linear relationship, $\lambda_{ij}$, is a measure of the covariation `pressure' on the pair $i,j$ due to the structural constraints. The pressure depends on {\em (i)} the difference between the contact maps of the native fold ($c_{ij}$) and the ones of the competing structures ($\bar c_{ij}$), see Fig.~4B; {\em (ii)} the proximity (in terms of energy, Eq.~[\ref{gap}]) of the native fold with its competitor structures; {\em (iii)} the strength of constraints in the design, controlled here by the inverse sampling temperature (Eq.~[\ref{slope}]). A consequence of {\em (i)} is that the intensities of the inferred couplings, $J_{ij}(a,b)$, measured through the scores $F_{ij}$ in Eq.~[\ref{APC}], are directly related to the local properties of the contact map on sites $i,j$ (and not on other sites), which explains the success of inverse approaches to disentangle direct from indirect effects in amino-acid correlations. However, the relationship also involves the average contact map $\bar {\bf c}$, originating from the competitor folds. As $\bar {\bf c}$, contrary to the native contact map $\bf c$, is not sparse, neither are the couplings $\bf J$. Couplings can, in particular, take non-zero values for $i,j$ in contact in the competing folds ($\bar c_{ij}$ close to 1), but not in the native fold ($c_{ij}=0$), see Figs.~4A\&B. Those `repulsive' couplings give rise to negative pressures $\lambda_{ij}$, and prevent the sequence from forming contacts that would result in the wrong fold, a clear illustration of negative design \cite{bere07,noivirt09,minning13}. Reciprocally, sites in contact in the native fold ($c_{ij}=1$) need not be subject to strong covariation and associated to strong couplings, if they are also in contact in the competing folds ($\bar c_{ij}$ close to 1). Our results clearly show that strong couplings do not necessarily imply, nor require structural contact (in the native fold). They also explain the success and limitations of contact prediction based on the magnitude of the inferred couplings (Fig.~2A).

A second major result of our approach, shown in Figs.~3A\&B, is that the inferred Potts model, when accurately inferred, here, with the ACE procedure, generates sequences that fold with high probability in the native state and show amino-acid diversity similar to the one observed in the original MSA. This performance is remarkable. While the energy of a LP sequence in a given structure is a sum of pairwise contributions only, the effective Hamiltonian of the sequences constrained to fold with high probability in this structure (and not in the competing folds) includes high-order multi-body interactions (originating from the competing folds) between amino acids, as can be expected for real proteins. Furthermore, independent-site models are unable to produce sequences with the correct fold (Fig.~3A). Our result corroborates the works of Ranganathan and collaborators \cite{Socolich:2005js,russ05}, who experimentally showed that sequences built according to a reshuffling of the MSA of the small WW domain respecting site conservation and (approximately) pairwise correlations folded correctly with a good (30\%) probability. The generative character of the Potts model, combined with the linear relationship between couplings and energetics parameters, agrees with previous studies showing that designed and real protein-like sequences are generated by Gibbs measure with Hamiltonian reflecting real protein energetics \cite{shakh93,bere07}. We stress that couplings  have to be inferred with great care, going beyond the mean-field DCA approximation, in order for the Potts model to be generative (Fig.~3A).

The analysis of real protein data could benefit from our analysis in several ways. First, the understanding of the relationship between the couplings and the contact map obtained here indicates that the pressures $\lambda_{ij}$ (Eq.~[\ref{scoring_projMJ}]) are excellent estimators of contacts (Fig.~2A), and may outperform the usual scores, $F^{(APC)}_{ij}$ (Eq.~[\ref{APC}]). The use of $\lambda_{ij}$ allows one to both keep track of the sign of the couplings (and avoid false positives corresponding to pressures originating from negative design, see above) and increase signal-to-noise ratio, by removing noise in the inferred couplings not aligned along the Miyazawa-Jernigan energy matrix. In real proteins, the physico-chemical properties of amino acids suggest the existence of different classes of contacts, whose energetics could be inferred from large-scale analysis of databases and used to define class templates $E$.  Projections of the inferred Potts couplings on those templates could in turn be used for contact predictions. Secondly, the detailed analysis of lattice proteins done here has allowed us to better understand how the existence of positive and negative design influences the covariation statistics of the MSA, and, hence, the properties of the inferred couplings. The relevance of negative design was already put forward in literature, {\em e.g.} in the study of the TIM-barrel fold protein \cite{bere07}. A promising path to further investigate negative design could be the study of coexistence between natively folded and misfolded proteins \cite{tuinstra08}. The direct estimate of ${\bf \bar c}$ in that case would be straightforward, as it would essentially correspond to the contact map of the misfolded competing structure. More generally $\bf \bar c$ in real proteins can be determined from the ensemble of misfolds obtained either from all-atom folding simulations \cite{Shaw2011} or from threading approaches. This is the subject of future work. Last of all, our work suggests a practical way to quantify the designability of a protein family, {\em i.e.} to measure the number of sequences `belonging' to the family. Current approaches rely on the computation of the maximal eigenvalue of the native contact matrix $\bf c$ \cite{england2003}, though our study suggests that the matrix $\bar {\bf c}$, gathering information about the competing protein folds, is also a determinant of the designability. Given the generative character of the inferred Potts model, a direct and more accurate measure of the designability is provided by its entropy computed with the ACE procedure, see Methods and Table~1. This approach can be applied to real protein families, an example is provided in \cite{entropy2015}.

In future studies, the total control we have over lattice proteins will allow us to study the importance of covariation and the success of inverse approaches in 'Gedanken' experiments related to many important issues of interest for real proteins, and going beyond the structural aspects studied here. Among those let us cite the detection of homology. Current protein databases, such as PFAM, classify query sequences into families based on Hidden Markov Model profiles, an extension of independent-site models capable of taking into account deletions and insertions \cite{eddybook}. It is an open problem to understand whether coupling-based models, exploiting covariation, could, contrary to HMM, recognize sequences with low homology in a reliable and computationally tractable way. Another very important issue is the estimation of fitness landscapes, or fitness changes in responses to one or more mutations \cite{Jacquier2013,Olson2014}. Covariation-based models have been recently introduced to predict escape paths of pathogens (virus or bacteria) from drugs or vaccines in this context \cite{Mann:2014ji}. Lattice proteins offer a unique benchmark to understand deeply and quantitatively and, ultimately, improve those approaches.      

\section*{Methods}

\vskip .5cm
\noindent{\bf Lattice-protein model}

\vskip .3cm \noindent
We consider model proteins, whose $L=27$ amino acids occupy the sites of  a $3 \times 3 \times 3$ cubic lattice \cite{gutin90,shakh93,li1996}. Four of the $103,406$ possible configurations of the backbone (excluding symmetries), hereafter called folds or structures, are shown in Fig.~2A. Unless otherwise said, we restrict ourselves to a representative subset of ${\cal N}=10,000$  folds \cite{heo11}. Two amino acids are said to be in  contact if they are nearest neighbors on the lattice (but not on the backbone). The contact map ${\bf c}^{(S)}$ of structure $S$ is its $27\times 27$ adjacency matrix: $c_{ij}^{(S)}=1$ if $i,j$ are in contact, 0 otherwise. Amino acids in contact interact through the Miyazawa-Jernigan (MJ) $20\times 20$ energy matrix $E$ \cite{jernigan85}.
The energy of a sequence ${\bf A} = \left( a_1, \cdots ,a_{27} \right)$ of amino acids folded into structure $S$ is
\begin{equation}
{\cal E}({\bf A}|S) = \sum_{i<j} c_{ij}^{(S)}\; E(a_i,a_j)\ .
\label{energy}
\end{equation}
The Boltzmann probability that sequence ${\bf A}$ folds into structure $S$ is given by (for unit temperature):
\begin{equation}
P_{\rm nat}(S|{\bf A}) = \frac{\displaystyle e^{-{\cal E}({\bf A},S)}}{\displaystyle \sum_{S'=1}^{\cal N} e^{-{\cal E}({\bf A},S')}}\ .
\label{pnat}
\end{equation}
Good folders are sequences $\bf A$ with large gaps between ${\cal E}({\bf A}|S)$ and the energies ${\cal E}({\bf A}|S')$ with the other folds \cite{shakh93}. 

\vskip .3cm
\noindent{\bf Sampling of sequence space}

\vskip .3cm \noindent
We generate a multi-sequence alignment (MSA) for the native fold, say, $S$, through Monte Carlo (MC) simulations with the Metropolis rule \cite{bere07}. The simulation starts from a sequence ${\bf A}$, and attempts to mutate randomly one amino acid at a time, say, $a_i\to a'_i$; Let ${\bf A}'$ denote the new sequence. If $P_{\rm nat}(S|{\bf A}')>P_{\rm nat}(S|{\bf A})$ the mutation is accepted; otherwise it is accepted with probability  $[P_{\rm nat}(S|{\bf A}')/P_{\rm nat}(S|{\bf A})]^{\beta}(<1)$. Parameter $\beta$ plays the role of an inverse algorithmic temperature that sets the stringency of sequence selection in the sampling MC procedure. The corresponding effective Hamiltonian, 
\begin{equation}\label{hlp}
{\cal H}^{\rm LP}[{\bf A}|S]=-\log P_{\rm nat}(S|{\bf A}) \ ,
\end{equation} 
includes not only the native-structure energy ${\cal E}({\bf A},S)$ in Eq.~[\ref{energy}], with pairwise interactions between the amino acids $a_i$, but also contributions coming from all the other folds $S'$ in Eq.~[\ref{pnat}], with multi-body interactions at any orders $\ge 2$.

In practice we choose $\beta=10^{3}$, which ensures that sequences fold in the native structure $S$ with probability $0.995$ or larger and that thermalization is fast (Fig.~A in S1 Text). A conservative estimate of the equilibration time of the MC dynamics (defined as the time needed for $P_{\rm nat}$ to relax to its equilibrium value independently of the initial sequence condition) is $<1,000$ MC steps. The MSA is then made of the sequences ${\bf A}^\tau$, with $\tau=1,\ldots, M= 50,000$, generated at regular intervals of 1,000 steps. Each sequence ${\bf A}^\tau$  is therefore drawn according to the equilibrium measure equal to $P_{\rm nat}(S|{\bf A}^\tau)^\beta$, up to a multiplicative normalisation factor. 

To mimic the biases in real protein MSAs, e.g. resulting from phylogeny, we may add to the effective Hamiltonian a penalty term proportional to the number of mutations with respect to a wild-type sequence ${\bf A}^{(wt)}$,
\begin{equation}
{\cal H}^{\rm LP}_{\rm biased}[{\bf A}|S,b] = -   \log P_{\rm  nat}(S|{\bf A}) + b  \sum_{i} \delta_{a_{i}, a^{(\rm wt)}_{i}} \ ,
\label{pdf_LP_biased}
\end{equation}
and repeat the MC sampling procedure above, with $M=100,000$. High values of the bias strength $b$ produce MSAs with high sequence identity, see Fig.~2B.

\vskip .3cm
\noindent{\bf Inference of Potts model from sequence data} 

\vskip .3cm \noindent
We compute the  one- and two-point amino-acid empirical frequencies computed from the sequences ${\bf A}^{\tau}$ in the MSA of a given native structure,
\begin{equation} \label{freq}
f_{i}(a) = \frac 1{M_{eff}} \sum_{\tau=1}^M \omega^{\tau} \delta_{a,a_i^\tau}  \ , \quad f_{ij}(a,b) = \frac 1{M_{eff}} \sum_{\tau=1}^M\omega^\tau \delta_{a,a_i^\tau}\, \delta_{b,a_j^\tau} \ ,
\end{equation}
where $\delta$ denotes the Kronecker function. The weight $\omega^{\tau}$  sequence ${\bf A}^\tau$ is defined as the inverse of the number of sequences in the MSA whose sequence identities with ${\bf A}^\tau$ are larger than $27\times(1-w)$, where $w$ is an arbitrary cut-off. The effective number of sequences is $M_{eff}= \sum_{\tau=1}^M\omega^\tau $. Throughout this work no reweighting was imposed ($w=0$ and $\omega^\tau=1$), except for the biased sampling case, where some results were obtained with cut-off $w=0.2$, see Fig.~2B.

We consider the maximum entropy (least constrained \cite{jaynes57a}) distribution over the sequences, that reproduces 
the one- and two-point empirical frequencies. This maximum entropy distribution is the Gibbs measure of a Potts model, with $q=20$-state variables, whose energy  is given by
\begin{equation} \label{Potts}
 {\cal H}^{\rm Potts}[{\bf A}; {\bf h},{\bf J}] = - \sum_{i} h_i(a_i) -  \sum_{i<j} J_{ij}(a_i,a_j)  \ ,
\end{equation}
where the sets of $L\times q$ fields, ${\bf h}=\{h_i(a)\}$, and of $L(L-1)/2 \times q^2$ couplings, ${\bf J}=\{J_{ij}(a,b)\}$, are chosen so that the frequencies computed from the Potts model distribution reproduce those of the data, Eq.~[\ref{freq}]. Note that, as the distribution depends on  the energies of the sequence configurations through their relative differences the Potts parameters are defined up to an arbitrary gauge (S1 Text, Section II). 

We resort to three methods to solve this hard computational problem. Within the mean-field Direct Coupling Approximation (DCA)  \cite{Morcos:2011c}, ${\bf J}$ is approximated as minus the pseudo-inverse of  the connected correlation matrix $\bf c$, with entries $c_{ij}(a,b) = f_{ij}(a,b) - f_i(a) f_j(b)$. Data are regularized through the introduction of a large pseudocount (S1 Text, Section II.B). The Pseudo-Likelihood Method (PLM) is a more sophisticated approximation to compute the direct couplings ${\bf J}$, exploiting the full MSA (and not only the frequencies). A large regularization $L_2$-penalty, $\gamma=0.01$, was considered during the inference \cite{ekeberg12} (S1 Text, Section II.C). The Adaptive Cluster Expansion (ACE) is an accurate but slower procedure, which recursively builds the most relevant clusters of strongly interacting sites \cite{Cocco:2011fo,Cocco:2012id,Bar:2015}. Couplings were inferred in the least-probable gauge, in which each interaction parameters corresponding to the least frequent amino acids are set to zero, with a weak $L_2$--penalty $\gamma =5/M$; $M$ is here the number of the sequences in the MSA. See S1 Text, Section II.D for more information about the implementation of the methods, and the ability of the Potts-ACE distribution to reproduce high-order statistics of the MSA.  In addition to ${\bf h}, {\bf J}$ ACE gives access to the entropy of the inferred Potts model, which is an estimate of the designability of the native fold.

For the simpler Independent-site Model (IM), the energy  $ {\cal H}^{\rm IM}$ is a sum of field contributions only (${\bf J}=0$ in Eq.~[\ref{Potts}]), see S1 Text, Section II.A for details.

\vskip 2cm
\noindent{\bf Contact prediction}  

\vskip .3cm \noindent
Computation of the Potts couplings $\bf J$ with DCA, PLM or ACE allows us to define the scores \cite{Morcos:2011c} (with the Averaged Product Correction \cite{Dunn08})
\begin{equation}
F^{(\rm APC)}_{ij} = F_{ij} - \frac{F_{i.} \, F_{j.}}{F_{..}}\ , \quad \text{where}\
F_{ij} = \sqrt{\sum_{a,b} J_{ij}(a,b) ^2} \ ,\label{APC}
\end{equation}
and  $F_{i.} = \sum _{l} F_{il}$, $F_{..} = \sum _{i} F_{i.}$. The values of the scores depend on the choice of the gauge. For the PLM and DCA procedures it is customary to choose the so-called zero-sum gauge,  in which the sums of the coupling matrix elements $J_{ij}(a,b)$ over each column or row vanish. This gauge is the one minimizing the Frobenius norm. For the ACE procedure, we have used the couplings inferred in the least-probable gauge, see paragraph above and S1 Text, Section II.D, to compute both the Frobenius (leading to black dots in Fig.~2A) and the projection (magenta symbols) scores; we have verified that contact predictions do not change for the four structures when scores are computed from couplings expressed in the zero-sum gauge. Within the mutual information approach, scores are simply given  by $MI_{ij} = \sum_{a,b} f_{ij}(a,b) \log [f_{ij}(a,b)/(f_i(a) f_j(b))]$ \cite{chiu91} (and are then subject to the above APC correction), and do not distinguish direct from indirect correlations. Scores are then ranked  in decreasing order, and used to predict contacts. The positive predictive value at rank $k$ is the fraction of the top $k$ scores whose corresponding pairs of sites $(i,j)$ are in contact.   

\vskip .3cm
\noindent{\bf Competitors, gap and average contact map}  

\vskip .3cm \noindent
We define the gap $\Delta(S|S_{nat})$ between structure $S$ and the native fold $S_{nat}$ through
\begin{equation}
e^{-\Delta (S|S_{nat})} = {\left\langle e^{- [ {\cal E}({\bf A}|S) - {\cal E}({\bf A}|S_{nat})]}\right\rangle}_{\bf A} \ ,
\label{gap}
\end{equation}
where the average is taken over sequences ${\bf A}$ in a MSA of $S_{nat}$. The distributions of gaps for  $S_{nat}=S_A,S_B,S_C,S_D$ are shown in Fig.~G in S1 Text. The closest competitor to $S_{nat}$ is defined as the structure $S$ minimizing $\Delta(S|S_{nat})$, see S1 Text, Section~III. We approximate the number $N_S$ of competitor folds and their typical gap $\Delta $ with the native structure $S_{nat}$ through $N_S \, e^{-\Delta} \equiv  \displaystyle{\sum_{S(\ne S_{nat})} e^{-\Delta (S|S_{nat})}}$. The average contact map of those competitors is defined as $\overline{c}_{ij} \equiv \frac 1{N_S \,e^{-\Delta}} \; \displaystyle{\sum_{S(\ne S_{nat})}e^{-\Delta (S|S_{nat})}\, c_{ij}^{(S)}}$.

\newpage
\vskip .8cm
%Figures

\noindent {\Large \bf Tables}

\vskip .6cm

\vskip 1cm

\begin{table}[h]
\begin{center}
\begin{tabular}{|c|c|c|c|c|}
\hline {\bf Structure} & {\bf Top eigenvalue}    & {\bf Entropy}  & {\bf Mean identity} \\
  & {\bf of $\bf c$}  &   {\bf (Potts-ACE)} & {\bf between seq. (\%)} \\
\hline $S_A$ & 2.5  &  49.5 & 23 \\
\hline $S_B$ & 2.6  & 47.5 & 24 \\
\hline $S_C$  & 2.9  & 57.4 & 19\\
\hline $S_D$  & 2.7 & 53.5 & 21\\
\hline
\end{tabular}
\end{center}
\label{tab:MSA caracteristic}
\end{table}%

\vskip .5cm
\noindent{\bf Table 1:  Designabilities and entropies of structures $S_A$ to $S_D$.}  

\noindent  Some estimates of how designable are the protein families associated to structures $S_A,S_B,S_C,S_D$:  largest eigenvalue of the contact map matrix $\bf c$ (1st column) \cite{england2003}, entropy of the Potts-ACE model (2nd column), shown to be generative in Fig.~3A, and mean percentage of identity between sequences in the attached multi-sequence alignments (MSA) (3rd column). The mean identity is defined as the number of sites carrying consensus amino acids, averaged over all sequences in the MSA and divided by the length of the protein ($L=27$); low identity corresponds to diverse MSA, and, hence, to large designability. According to our estimates of the entropies, the volumes (Fig.~1) associated to structures $S_B$ and $S_C$ are, respectively,  of about $4\,10^{20}$ and $8.5\,10^{24}$ sequences, while the total number of sequences is $20^{27}\simeq 10^{35}$. For more information about the meanings of designability and entropy, see Section I.B in Text S1. 

\vskip 2.5cm

\begin{table}[h]
\begin{center}
\begin{tabular}{|c|c|c|c|c|c|c|}
\hline {\bf Structure} & { \bf MI} &  {\bf  DCA } & { \bf PLM}  & {\bf  ACE } & {\bf Projection} & {\bf Average pressures}  \\
\hline $S_A$   & 20 & 19 & 24 & 24 & 27 & 2.00\\
\hline $S_B$  & 19 & 20 & 24 & 24 & 27 & 2.09 \\
\hline $S_C$ & 19 & 21 &  25 & 24& 27 & 1.32\\
\hline $S_D$   & 16 & 19 & 22 & 25 & 27 &  1.78 \\
\hline
\end{tabular}
\end{center}
\end{table}

\vskip .5cm
\noindent{\bf Table 2:  Quality of contact prediction with the different methods of inference, and average pressure.}  

\noindent  Number of correctly predicted contacts after 28 predictions with the methods MI, DCA, PLM, ACE, and Projection,  see Fig.~2A,
and pressures $\lambda$, defined in Eq.~[\ref{scoring_projMJ}], averaged over the pairs of amino acids in contact in the native folds.

\newpage

\section*{Supporting Information Files}

{\bf Text S1.}
Supporting information for Benchmarking inverse statistical approaches for protein structure and design with exactly solvable models. (PDF)

\vskip .5cm
\noindent{\bf Code S1.}
Multi-sequence alignments,  inferred couplings, list of lattice proteins structures, codes to generate lattice proteins and compute probabilities of folding. See readme.txt file. (ZIP)

\vskip .5cm
\noindent{\bf Code S2.}
Files for inference with ACE procedure for structures $S_A$ and $S_B$. See readme.txt file. (ZIP)

\vskip .5cm
\noindent{\bf Code S3.}
Files for inference with ACE procedure for structures $S_C$ and $S_D$. See readme.txt file. (ZIP)

\section*{Acknowledgments}
This work was initiated during an internship of G. R\'emy, to whom we are grateful. We also thank M. Weigt for many useful discussions, and A. Coucke for her help in the analysis of contact predictions and MJ projections. 
% le
%\nolinenumbers

\vskip 1cm

\noindent {\Large \bf Figure captions}

\vskip 1cm

{\bf Figure 1 : From lattice-protein sequence space to inferred Potts model.} 
Protein families, each corresponding to a particular structure $S$, represent portions of sequence space (colored blobs), in which all sequences (colored dots) fold into a unique conformation. Many sequences are expected to be non folding, and not to belong to any family (black dots). Protein families differ by how much they are designable, {\em i.e.} by the numbers of sequences folding onto their corresponding structures, represented here by the sizes of the circles.  $S_A$ and $S_B$ are the least designable folds, while $S_C$ and $S_D$ are realized by larger numbers of sequences, see Table~1. From a multi-sequence alignment (MSA) of one family, we infer the maximum-entropy pairwise Potts model reproducing the low-order statistics of the MSA. The model is then tested for structural prediction and generating new sequences with the same fold. An important issue is to unveil the meaning of the inferred pairwise couplings $\bf J$, which depend both on the family fold, as well as on the competitor folds.

\vskip 1cm

{\bf Figure 2 : Inverse statistical approaches are able to extract structural information from sequence data of lattice proteins.}
{\bf A.}  {\em Left:} Structures $S_A,S_B,S_C,S_D$. Amino acids (blue circles) are shown with their number, running from 1 to 27 along the protein backbone (black line). There are $28$ contacts between nearest-neighbor amino acids not supported by the backbone.  
{\em Right:} Positive predictive values (PPV), defined as the fraction of contacts among the $k$ top scores, with the MI, DCA, PLM, ACE, and Projection procedures (Methods). Multi-sequence alignments with $M=5 \cdot 10^4$ sequences were generated  with Monte Carlo sampling at inverse temperature $\beta=10^{3}$(Methods).  
{\bf B}. Structural predictions for biased alignments ($M=10^5$ sequences). The left panel shows the PPV with the PLM procedure for MSA for structure $S_B$, generated with four values of the bias $b$ (Methods). Squares and dots correspond to predictions done with and without reweighting respectively (Methods). Predictions for the weakest bias ($b=0.01$) are identical to the unbiased case ($b=0$) shown in Fig.~2A. The right panel reports the histograms of the Hamming distances to the Wild Type (WT) sequence, KCLIDRTEFKAREVLVPAKCCEFKECL, randomly chosen among the unbiased MSA of $S_B$. The effective number of sequences  (Methods) in the MSAs where $M_{eff}=100000, 75000, 6378, 102$ for, respectively, $b=0.01, 0.05, 0.075, 0.1$.

\vskip 1cm

{\bf Figure 3 : Inferred Potts-ACE model generates sequences with high folding probabilities and diversities.}
{\bf A.}  Folding probabilities $P_{nat} (S_B|{\bf A})$, Eq.~[\ref{pnat}], for four sets of $10^4$ sequences $\bf A$ randomly generated with the Independent-site Model (IM, green), the Potts-ACE (red), the Potts-PLM (orange) and the Potts-Gaussian (blue) models vs. their Hamming distances to the consensus sequence of the `natural' MSA of structure $S_B$ used to infer the four models. Black symbols show results for the `natural' sequences, sampled at inverse temperature $\beta=10^3$ (Methods). Most sequences drawn from the Potts-ACE model have high folding probabilities, while most sequences drawn from the IM have low values of $P_{nat}$; sequences generated with the Potts-PLM model lie somewhere in between. Sequences drawn from the Potts-Gaussian model have very high folding probabilities, but are very close to the consensus sequence, and fail to reproduce the diversity of sequences seen in  the `natural' MSA (black) and Potts-ACE (red) data. Hamming distances for the Potts-ACE and PLM models have been shifted by, respectively, $+\frac 14$ and $-\frac 14$ to improve visibility. Filled ellipses show domains corresponding to one standard deviation of the effective Hamiltonian $\beta {\cal H}^{\rm LP}[{\bf A}|S]=-\beta \log P_{nat}(S|{\bf A})$, Eq.~[\ref{hlp}]. 
{\bf B}. Scatter plot of  the `energy' ${\cal H}^{Potts-ACE}[{\bf A}; {\bf h},{\bf J}]$, Eq.~[\ref{Potts}], with the inferred Potts-ACE ($x$-axis) vs. effective Hamiltonian $\beta {\cal H}^{\rm LP}[{\bf A}|S]=-\beta \log P_{nat}(S|{\bf A})$, Eq.~[\ref{hlp}] ($y$-axis), for the sequences in the MSA generated with the Potts-ACE model for structure $S_B$. Only sequences within the 90\%-100\% percentiles of  $P_{nat}$ values are shown. Colors identify intervals of values for $P_{nat}$, see legend in panel. The energy of the sequences computed with the Potts-ACE model have been subtracted the energy of the best folder, such that the minimal energy is zero.

\vskip 1cm

{\bf Figure 4 : Inferred Potts couplings encode energetics and structural information about native and competitor folds, reflecting both positive and negative designs.}
{\bf A.} Values of $J_{ij}(a,b)$ (inferred from a MSA of structure $S_B$ with the Potts-ACE method) vs.  $-E(a,b)$  across all pairs of sites $i,j$ and of amino acids $a,b$ (found at least once in the MSA on those sites). Couplings and MJ energy parameters are shown in  the consensus gauge, in which the entries attached to the most probable amino acids in each site are fixed to zero. Red symbols correspond to pairs $(i,j)$ in contact, while blue symbols correspond to no contact. 
{\bf B.} Lower-triangle: contact map  $c_{ij}$ of structure $S_B$. Full blue squares correspond to pairs of sites $i,j$ in contacts. Green and red dots show, respectively,  true and false positives among the 28 largest scores $F^{APC}_{ij}$ with the ACE method (Methods). Upper triangle: average contact map $ \bar c_{ij}$, computed over all competitor folds weighted with their Boltzmann weights (Methods). The four missed contacts (all touching the central site 4) correspond to large $\bar c_{ij}$. Red squares locate the four false positives.
{\bf C.} Pressure $\lambda_{ij}$ for each pair of sites $(i,j)$, computed from Eq.~[\ref{scoring_projMJ}], vs. $c_{ij} - \overline{c}_{ij}$ for structure $S_B$.  The 195 pairs of sites which can never be in contact on any fold  due to the lattice geometry are shown with magenta pluses. The 28 contacts on $S_B$ (red symbols) are partitioned into the Unique-Native (UN, 14 full triangles) and Shared-Native (SN, 14 empty triangles) classes, according to, respectively, their absence or presence in the closest competitor structure, $S_F$ (Fig. 4D).  The remaining 128 pairs of sites (blue symbols) are not in contact on $S_B$, and are partitioned into the Closest-Competitor  (CC, 14 full squares) and the Non-Native (NN,  114 empty squares) classes, according to, respectively, whether they are in contact or not in the closest competitor structure, $S_F$.  Similar results are found for  $S_A$, $S_C$ and $S_D$, see Table~2 and Figs.~H, I \& J in S1 Text. As in Fig.~4A, we use coupling and MJ entries expressed in the consensus gauge, since the consensus sequence corresponds, or is close to the best folding sequence, used as a reference sequence in our theoretical calculation of the pressure (S1 Text, Section III). Changing the gauge {\em e.g.} to the least-probable gauge affects the amplitudes of the pressures $\lambda_{ij}$, but does not qualitatively alter the results.
{\bf D}. Structure $S_F$, the closest competitor structure to $S_B$. Note that the four missed contacts (among the top 28 $F^{APC}$ scores with the ACE method) are carried by the center of the cube (site $i=4$ on $S_B$ and $S_F$), see fold $S_B$ in Fig.~2A and its contact map in Fig.~4B. Two of the four false positives are contacts on $S_F$, and are thus in the CC class.

\newpage

\noindent {\Large \bf Figures}

\vskip 1cm

\begin{figure}[h!]
\includegraphics[width=16cm]{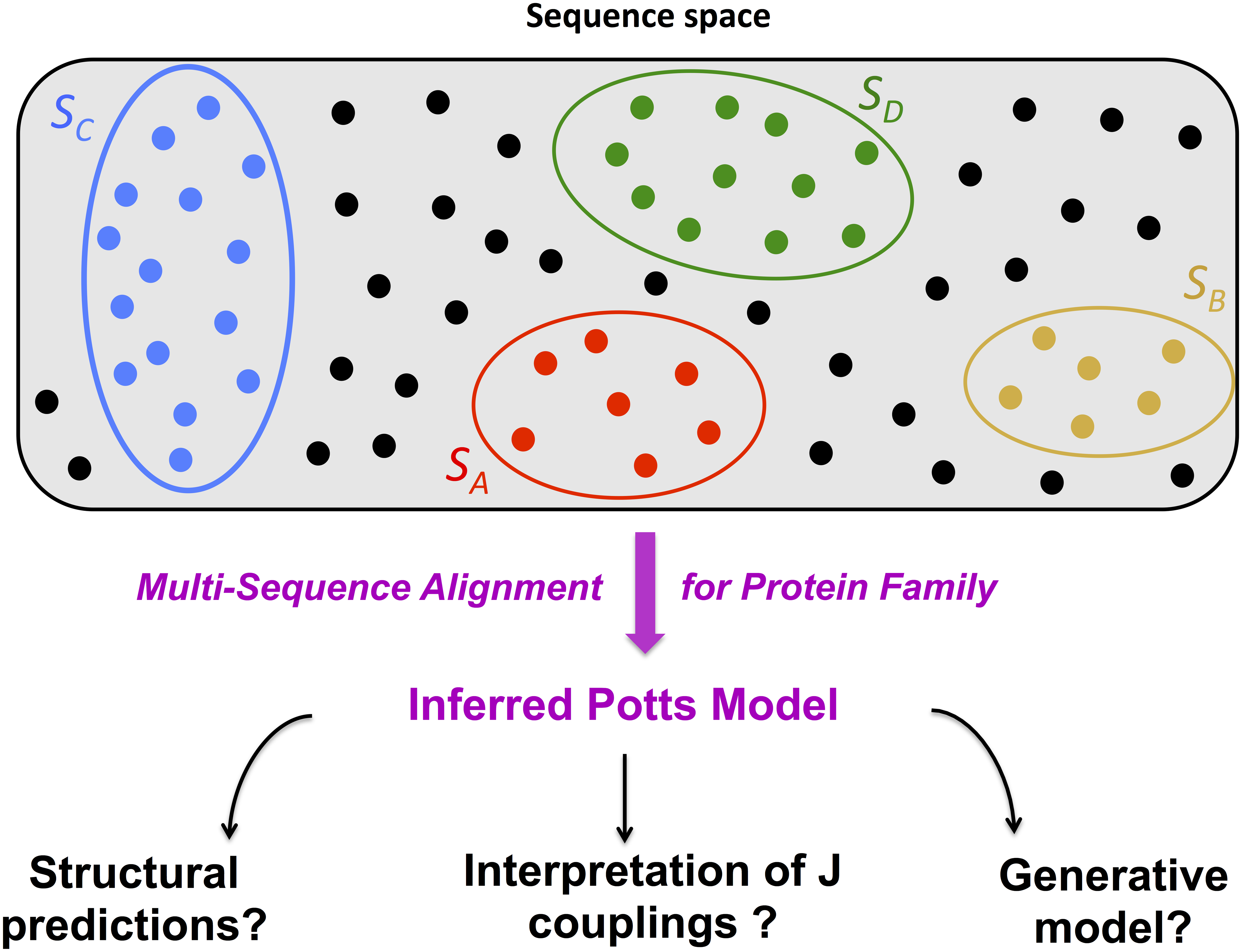}
\caption{}
\end{figure}

\newpage

\begin{figure}[h!]
\includegraphics[width=14cm]{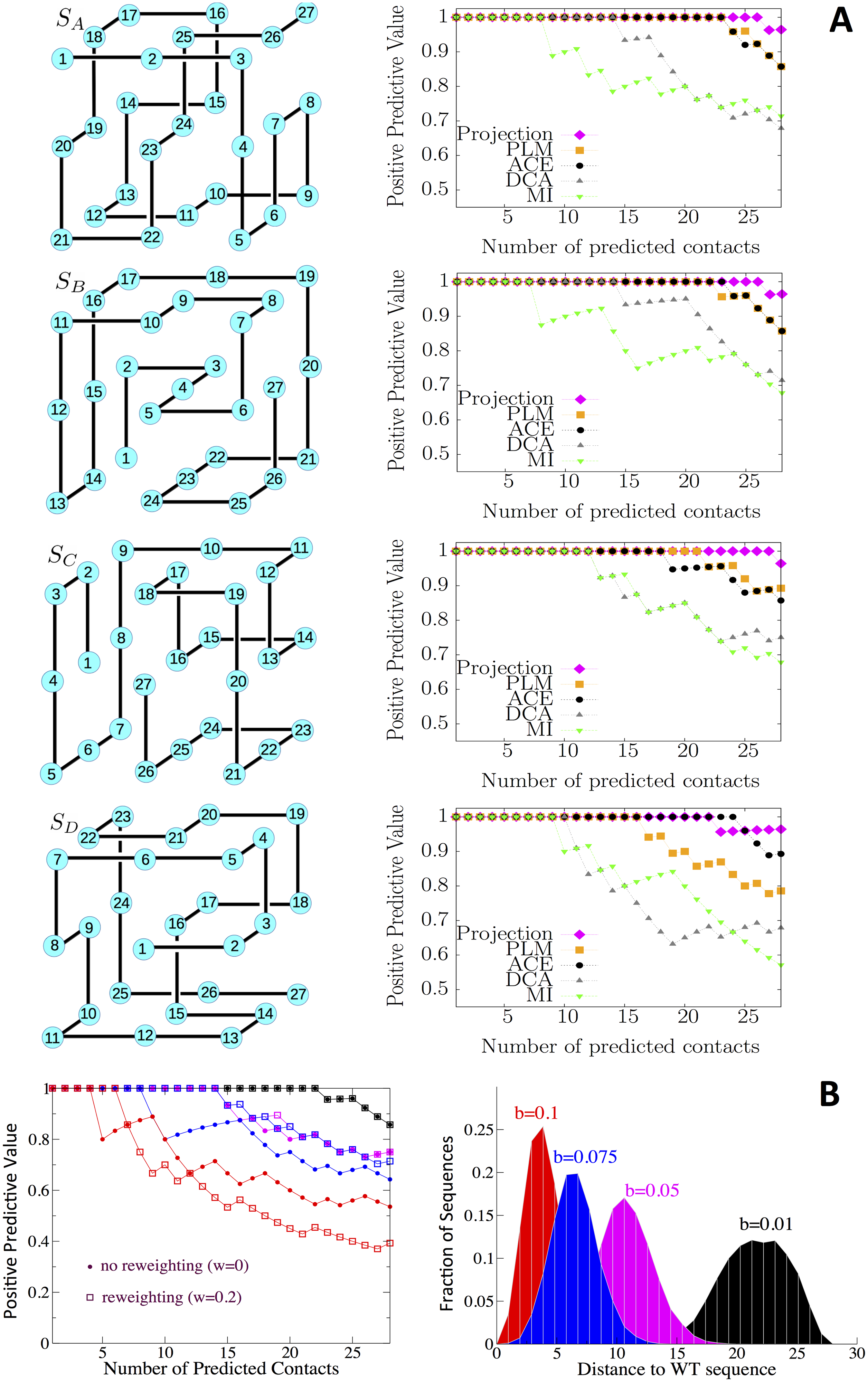}
\caption{}
\end{figure}

\newpage

\begin{figure}[h!]
\includegraphics[width=14cm]{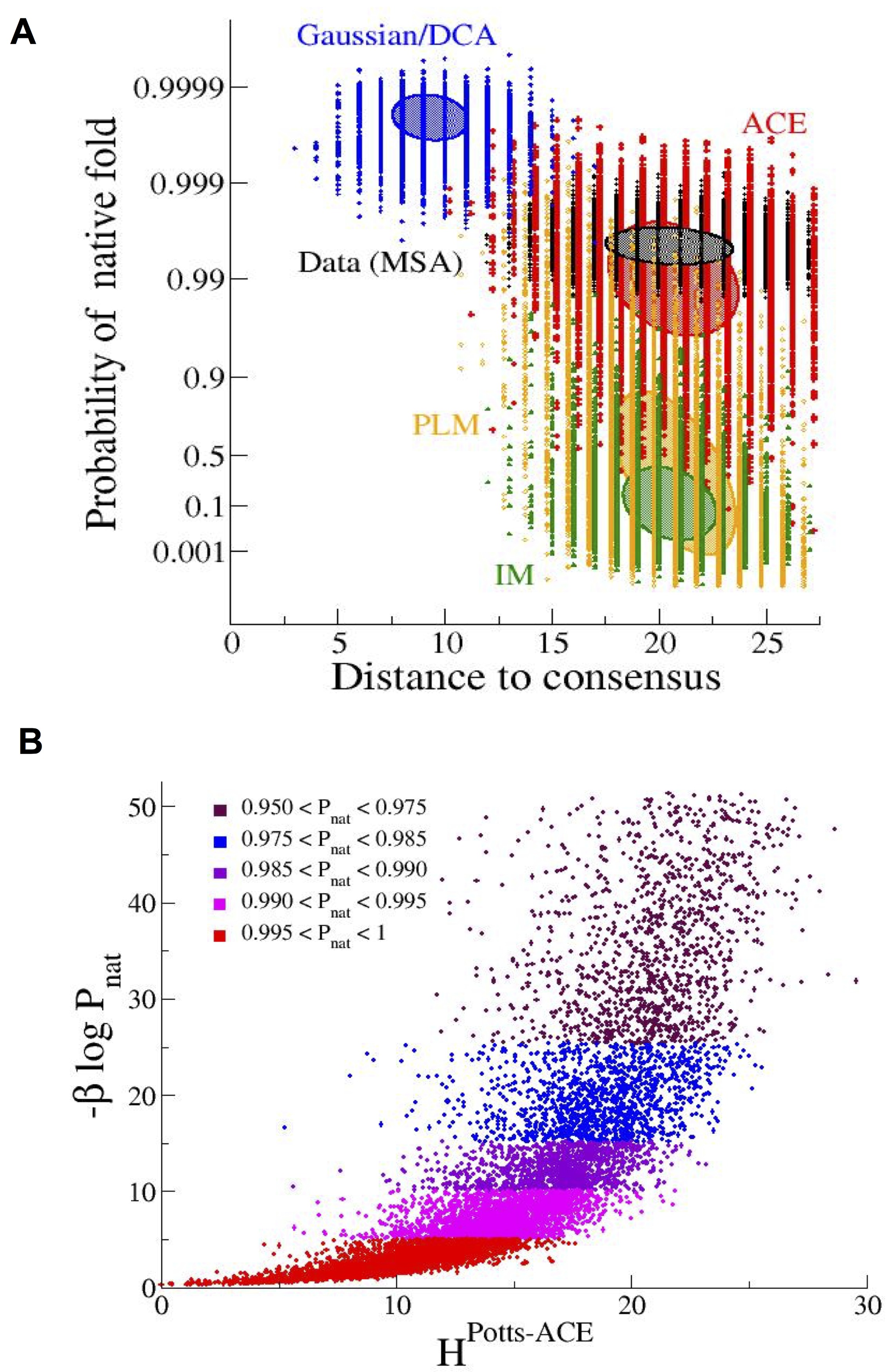}
\caption{}
\end{figure}

\newpage

\begin{figure}[h!]
\includegraphics[width=16cm]{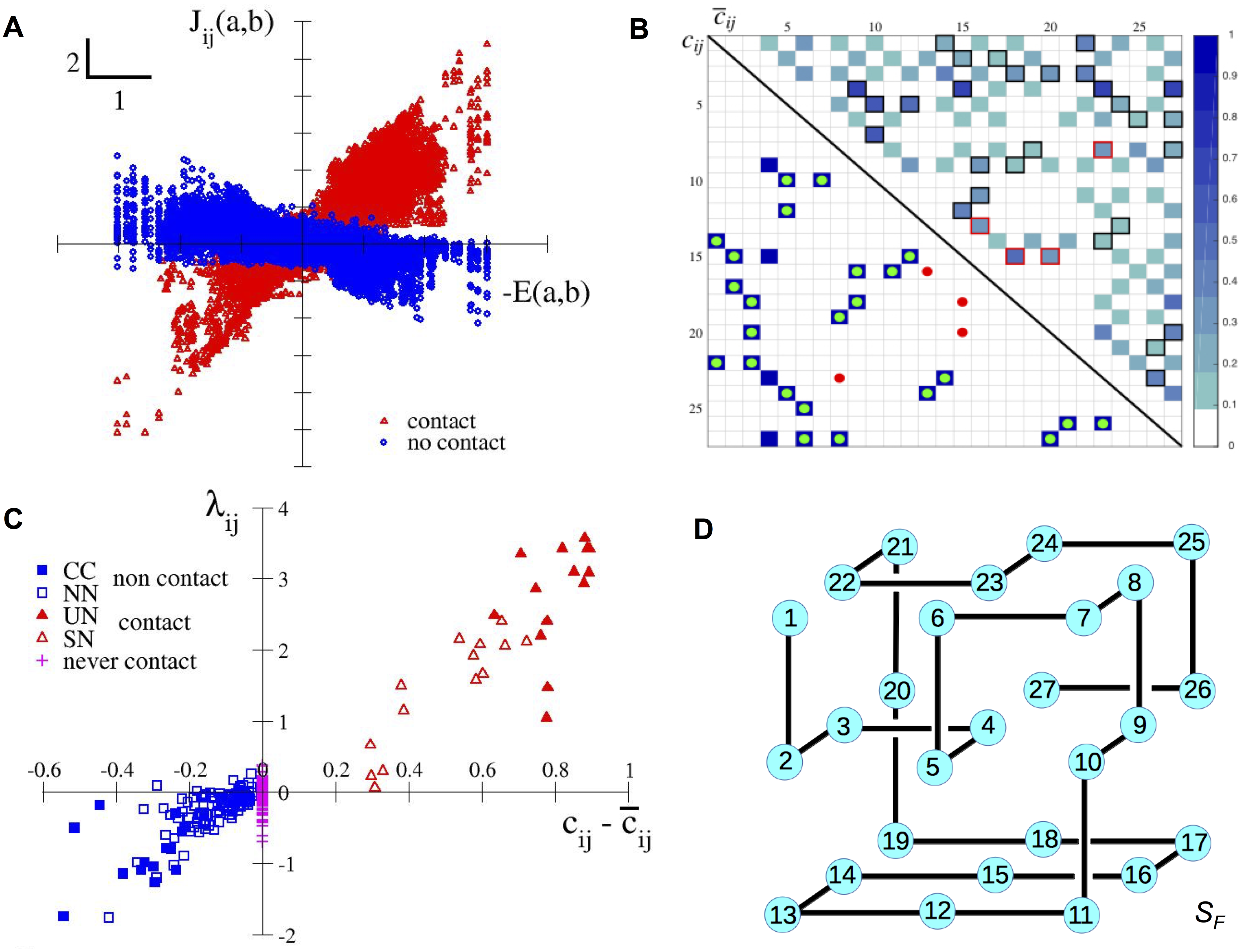}
\caption{}
\end{figure}

\end{document}